\documentclass{article}
\usepackage[T1]{fontenc}
\usepackage{spconf,amsmath,graphicx}
\usepackage{adjustbox}
\usepackage{cite}
\usepackage{hyperref}

\usepackage{caption}
\usepackage{amsfonts}
\usepackage{booktabs}
\title{SRP-PHAT-NET: A Reliability-Driven DNN for Reverberant Speaker Localization}
\name{Bar Shaybet\textsuperscript{1}, Vladimir Tourbabin\textsuperscript{2} and Boaz Rafaely\textsuperscript{1}}
\address{\textsuperscript{1}School of Electrical and Computer Engineering, Ben-Gurion University of the Negev
\\
\textsuperscript{2} Reality Labs Research @ Meta
}

\begin{document}
%\ninept
%
\maketitle
\begin{abstract}
Accurate Direction-of-Arrival (DOA) estimation in reverberant environments remains a fundamental challenge for spatial audio applications. While deep learning methods have shown strong performance in such conditions, they typically lack a mechanism to assess the reliability of their predictions—an essential feature for real-world deployment. In this work, we present the SRP-PHAT-NET, a deep neural network framework that leverages SRP-PHAT directional maps as spatial features and introduces a built-in reliability estimation. To enable meaningful reliability scoring, the model is trained using Gaussian-weighted labels centered around the true direction. We systematically analyze the influence of label smoothing on accuracy and reliability, demonstrating that the choice of Gaussian kernel width can be tuned to application-specific requirements. Experimental results show that selectively using high-confidence predictions yields significantly improved localization accuracy, highlighting the practical benefits of integrating reliability into deep learning-based DOA estimation.
\end{abstract}
\section{Introduction}
\label{sec:intro}

Direction-of-Arrival (DOA) estimation plays a fundamental role in numerous applications across diverse domains, including teleconferencing, hearing aids, robotics \cite{roboticsAudio,roboticsAudio2} and acoustic surveillance. More recently, the advent of virtual reality (VR) and immersive multimedia has further underscored the importance of precise DOA estimation for generating realistic, spatially accurate soundscapes\cite{doa_vr}.
These VR and other speech-centric applications require accurate speaker localization, identifying the direction of one or more talkers in enclosed environments such as meeting rooms and auditoria. DOA estimation in such environments becomes a challenge due to background noise and reverberation, which must be addressed to ensure reliable and effective localization performance.

Over the past few decades, numerous algorithms have been proposed for DOA estimation. Classical examples include beamforming-based methods such as Steered Response Power with Phase Transform (SRP-PHAT), and subspace-based approaches like Multiple Signal Classification (MUSIC) \cite{music,srp2} and Estimation of Signal Parameters via Rotational Invariance Techniques (ESPRIT) \cite{ESPRIT}. These methods have shown strong performance in controlled settings; however, they often struggle in real-world environments where reverberation and background noise introduce significant challenges.

Consequently, there has been growing interest in employing data-driven approaches, especially deep learning, to overcome these limitations and achieve more robust performance under realistic acoustic conditions.
Deep neural networks (DNNs) have shown remarkable success in many tasks related to speech and audio processing, including enhancement \cite{ambi_algo_speach_enhancment}, source separation \cite{ambi_algo_source_seperation_DNN}, and speaker localization\cite{DOA_CNN}. By leveraging large training datasets and powerful network architectures, deep learning methods outperform traditional, model-based algorithms in adverse acoustic conditions, showing higher resilience to noise and reverberation. However, the development of deep learning-based DOA estimators introduces its own set of challenges such as the need for extensive and diverse training data to cover the many possible acoustic scenarios that might be encountered in real-world settings \cite{shaybet}.

A key aspect of practical DOA estimation systems is not only achieving high accuracy but also understanding the reliability of each prediction. In this context, reliability refers to the model’s ability to produce a confidence score that reflects the true accuracy of its estimates. Such scores are essential for informed decision-making, enabling systems to dynamically adjust processing strategies depending on the required trade-off between speed and certainty. For example, real-time applications in dynamic environments may prioritize fast decisions with moderate confidence, while static scenarios involving relatively stationary speakers can afford longer processing times to improve accuracy. Some traditional model-based methods have addressed reliability explicitly. The Direct-Path Dominance (DPD) test \cite{DPD_Nadiri,DPD_orel}, for instance, identifies time-frequency bins that are more likely to represent direct-path signals and thus yield more accurate DOA estimates, particularly in reverberant environments. In tracking scenarios, the Kalman filter \cite{kalmanFilter} inherently estimates the reliability of measurements through its internal covariance calculations, weighting each observation according to its confidence. Despite the impressive performance of deep learning methods, recent DNN-based DOA estimators typically lack such built-in reliability measures. They often output predictions without any indication of their confidence \cite{DOA_CNN, doa_vr, doa_deep2, doa_deepMUSIC, doa_crnn, doa_deep, doa_dnn_cnn, doa_dnn_seldnet}, limiting their applicability in real-world systems that require robustness and adaptability. While uncertainty estimation in neural networks is an active area of research \cite{UQ_stt}, these methods have not yet seen widespread adoption in DOA applications and the important feature of reliability does not seem to be available in recent DNN methods for DOA estimation.

In this work, we address the lack of reliability estimation in deep learning-based DOA systems by introducing a model that not only outperforms classical approaches such as SRP-PHAT under reverberant conditions, but also provides a built-in confidence score that reflects the reliability of its predictions—a critical feature for real-world deployment.

Specifically, we propose using the model output value of the predicted class as a proxy for reliability. To enable this, the model is trained using Gaussian-weighted labels centered around the ground-truth direction, which encourages the network to learn smoother and more calibrated confidence estimates. We further analyze the impact of this training strategy on both the model’s accuracy and reliability, and demonstrate that selecting the appropriate Gaussian kernel width can be guided by the target reliability requirements.

\noindent
The contributions of this work are as follows:
\begin{itemize}
  
  \item The development of a deep neural network for DOA estimation that not only produces accurate predictions in adverse acoustic conditions (as do similar networks), but also outputs a reliability score based on the output score. We demonstrate that this novel additions can significantly improve DOA estimation accuracy by using estimates with higher reliability. This approach, while demonstrated for a specific network, could potentially be applied to other DOA estimation networks that are based on classification.
  
  \item The proposal of training the model using a Gaussian-weighted labeling and the systematic analysis of the impact of different training methodologies and gaussian kernel widths on both the accuracy and reliability of DOA predictions. Specifically, we show that the choice of kernel can be tailored to the application, depending on whether the priority is to obtain many estimates of any quality or fewer estimates of higher quality.

\end{itemize}

% The remainder of this paper is organized as follows. (??? should be completed)
% We first survey related work in model-based and deep learning-based DOA estimation, highlighting the existing solutions and their limitations with respect to array geometry variations. Next, we describe the theoretical underpinnings of SAP-PHAT features and illustrate how they can serve as robust input representations for DOA estimation. We then detail the design of our proposed DNN architecture and training methodology, followed by an extensive evaluation on multiple array configurations and acoustic scenarios. Finally, we discuss insights, future directions, and the broader implications of our findings for spatial audio and emerging technologies.

\begin{figure*}[t]
    \centering
    \includegraphics[width=\textwidth]{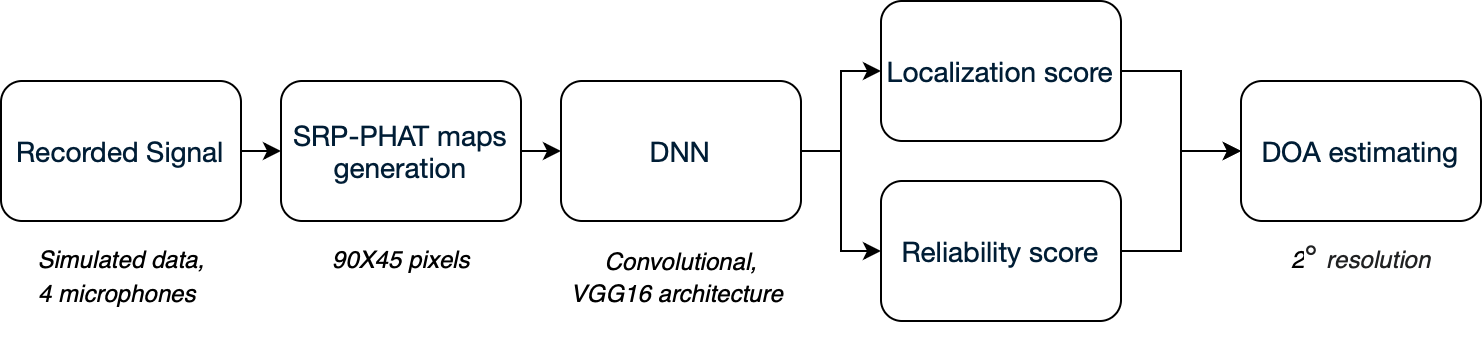} 
    \caption{Overview of the algorithm pipeline.}
    \label{fig:algorithmDiagram}
\end{figure*}

\section{SRP-PHAT overview}
\label{sec:SRP_PHAT_overview}
This section presents an overview of the method used to extract directional map from the array signals.
\subsection{SRP model}
\label{sec:SRP_model}

Steered Response Power (SRP) is a method that computes the energy of signals received by a microphone array to estimate the location of a sound source. The core idea involves steering the array’s response to different candidate locations in space by applying calculated time delays to the signals from each microphone. The algorithm operates on the principle of constructive interference: when the time delays align the signals correctly for the true source location, the algorithm exhibits maximum energy. By scanning a spatial grid of candidate locations, the algorithm identifies the point with the highest steered response power as the estimated source position.

For an array of M microphones, the steered response power $P(\theta,\phi)$ for a candidate source direction $\theta,\phi$ where $\theta$ is the elevation angle [0,180] and $\phi$ is the azimuth angle [0,360) is formulated as:

\begin{equation} \label{eq:p_steering}
P(\theta,\phi) = \sum_{m=1}^{M-1} \sum_{n=m+1}^{M} 
R_{m,n}\bigl(\tau_{m,n}(\theta,\phi)\bigr)
\end{equation}

\noindent
where $R_{m,n}(\cdot)$ is the cross-correlation of the signals received at microphones $m$ and $n$ and $\tau$ is the time difference of arrival (TDOA) for the hypothesized direction $(\theta,\phi)$.

\subsection{SRP-PHAT model}
\label{sec:SRP_phat_model}
SRP-PHAT extends the basic SRP approach, the key enhancement lies in applying a Phase Transform (PHAT) weighting to the cross-correlation function. The cross correlation is described as follow between microphone $m$ and $n$:

\begin{equation}\label{eq:cross_correlation}
R_{X_m X_n}(\tau) 
= \int_{-\infty}^{\infty}
\psi_{mn}(\omega)\, X_m(\omega)\, X_n(\omega)\, e^{j \omega \tau}\, d\omega
\end{equation}

\noindent
where $\psi_{mn}$ is the weighting function. The PHAT weight function is expressed as: 

\begin{equation}\label{eq:phat}
\psi_{\mathrm{PHAT}}(\omega) 
= \frac{1}{\bigl|X_m(\omega)\,X_n^*(\omega)\bigr|}
\end{equation}

\noindent
This weighting function normalizes the magnitude of the signals, thereby emphasizing the phase information, which is less affected by noise and environmental distortions. The following formula described the cross-correlation with the PHAT weighting function, substituting 
\eqref{eq:phat} into \eqref{eq:cross_correlation}:

\begin{equation}
\label{eq:crosscorr_withphat}
R_{X_m X_n}(\tau) 
= \int_{-\infty}^{\infty} 
\frac{X_m(\omega)\,X_n(\omega)}
{\bigl|X_m(\omega)\,X_n^*(\omega)\bigr|}
e^{j \omega \tau}\, d\omega
\end{equation}

\noindent
In our research the PHAT weighting function was chosen over other functions such ML, ROTH and SCOT \cite{normal_ROTH,normal_ML_PHAT} due to its popularity.
SRP-PHAT results with directional map where each pixel is the power correspond to spatial angle $(\theta,\phi)$ as described in Eq. \ref{eq:p_steering}, where x axis is the azimuth angle $\phi$ and y axis is the elevation angle $\theta$. As seen, the higher value in the directional map implies for source from that direction. 

By Eq. \ref{eq:p_steering} it can be notice that the map is assembled from cross-correlation between pairs of microphones. As more microphone available, more pairs exists and the ambiguity reduced. So the classic SRP-PHAT map algorithm determines the source location by choosing the direction with the highest value as follows:
\begin{equation}
(\phi,\theta)_s = \arg\max_{\phi,\theta} \bigl(P\bigr)
\end{equation}

\noindent
where $(\phi,\theta)_s$ is the source estimated direction.

\section{SRP-PHAT-NET algorithm} \label{sec:algorithm_overview}

To provide a clear understanding of the proposed SRP-PHAT-NET framework, a high-level overview of its algorithmic structure is presented. The first stage involves extracting directional maps from multichannel audio recordings using the SRP-PHAT method, producing spatial features. In the second stage, the directional maps are preprocessed to standardize input format and enhance compatibility with the neural network. The third stage employs a deep neural network based on the VGG-16 architecture \cite{vgg_original} to classify the direction of arrival (DOA) across discrete azimuthal bins. The fourth and final stage derives a reliability score from the softmax output of the network, providing a confidence estimate for each prediction. A block diagram illustrating the complete processing pipeline is shown in Fig.\ref{fig:algorithmDiagram} The following subsections describe each stage in detail.

\subsection{Overview of SRP-PHAT networks} 
\label{sec:overview_srp_networks}

Several recent studies have leveraged SRP-PHAT directional maps as input features for deep-learning architectures, utilizing their robustness to reverberation and noise while benefiting from neural networks' ability to resolve spatial ambiguities and improve localization accuracy. Some work focused on performance improvement over classical SRP-PHAT in various acoustic conditions using fully connected or CNN architectures \cite{Direction_of_Arrival_Estimation_with_Microphone_Arrays_Using_SRP-PHAT_and_Neural_Networks,Sound_Source_Localization_Based_on_SRP-PHAT_Spatial_Spectrum_and_Deep}. A causal three-dimensional CNN framework was introduced that processes sequences of SRP-PHAT maps, enabling accurate tracking of moving sound sources in challenging acoustic scenarios \cite{Robust_Sound_Source_Tracking_Using_SRP-PHAT_and_3D_Convolutional}. Several studies leveraged the spherical representation of SRP-PHAT maps to develop spherical convolutional recurrent neural networks, enabling efficient DOA estimation with reduced computational complexity\cite{SPHERICAL_CONVOLUTIONAL_RECURRENT_NEURAL_NETWORK_FOR_REAL-TIME_SOUND_SOURCE_TRACKING,DOA_Estimation_for_Spherical_Microphone_Array,DeepSphere}. Different sampling techniques for SRP-PHAT maps have also been explored, including the use of icosahedral CNNs applied directly to SRP-PHAT maps, leveraging spherical geometry to achieve robust direction-of-arrival estimation with 3D microphone arrays \cite{DOA_Estimation_of_Sound_Sources_Using_Icosahedral_CNNs}. Collectively, these methods highlight a growing consensus on the value of SRP-PHAT directional maps as input features, demonstrating substantial improvements over classical SRP-PHAT across diverse acoustic scenarios and providing robust localization and tracking capabilities. 

The algorithm proposed in this work shares similarities with several previously developed SRP-PHAT-based deep learning approaches, particularly in utilizing convolutional neural networks to process directional maps for accurate DOA estimation \cite{Robust_Sound_Source_Tracking_Using_SRP-PHAT_and_3D_Convolutional,SPHERICAL_CONVOLUTIONAL_RECURRENT_NEURAL_NETWORK_FOR_REAL-TIME_SOUND_SOURCE_TRACKING}. However, prior studies consistently estimates DOA used for all frames, without considering the reliability of these estimates, which may vary under the condition of reverberation, for example. Our work explicitly addresses this gap by integrating a reliability estimation mechanism into the model, providing a confidence score alongside each prediction, showing that this additional information can improve performance significantly.

In the following sections, we provide a detailed explanation of the algorithm implementation. This includes the process of extracting SRP-PHAT directional maps as input features, the design of the deep neural network architecture used for direction-of-arrival estimation, and the integration of a reliability estimation mechanism. Each component is described in terms of its role within the overall system, highlighting how the model is structured not only to support accurate predictions but also to quantify the confidence of those predictions through a dedicated reliability score.

\subsection{Directional Maps Generation} \label{sec:srp:extracting_directional_map}
   
The process of generating SRP-PHAT direction maps begins with the definition of the spatial sampling grid over which the source location is estimated. Various angular grid configurations have been explored in prior works \cite{srp_spherical_conv,doa_srp_icosehedral}, however, for simplicity, a uniform grid with 1° resolution in both azimuth and elevation is adopted, forming a 360×90 grid. Given a multichannel input signal from a microphone array, the SRP-PHAT energy at each grid point is calculated using the formulation in Eq. (\ref{eq:p_steering}). This procedure yields a scalar energy value for each direction, resulting in a full SRP-PHAT map that reflects the spatial likelihood of source presence. Although the final input to the neural network is a lower-resolution 90×45 map, we first compute the high-resolution version to capture fine spatial details and avoid artifacts or loss of directional information. Downsampling is applied only after the full map is constructed, ensuring that the reduced representation still preserves the essential spatial structure. Finally, the map values are normalized to a fixed numerical range.

\subsection{DNN model} \label{sec:srp:DNN_model}
\begin{figure*}[t]
    \centering
    \includegraphics[width=\textwidth]{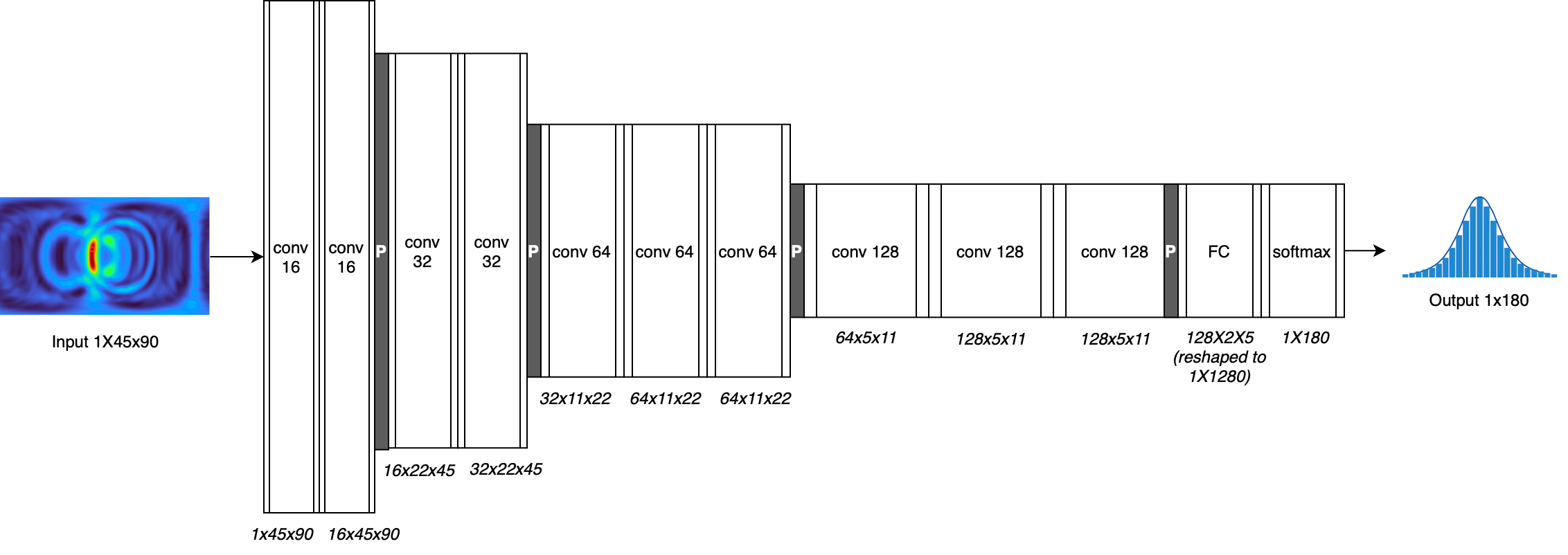} 
    \caption{Model architecture of the proposed algorithm. Numbers in the layers indicates output channels, the number under the layers indicates the input size.  }
    \label{fig:model_architecture}
\end{figure*}

A DNN model based on the VGG-16 architecture \cite{vgg_original} was selected as the primary component; its architecture is detailed in Fig. \ref{fig:model_architecture}. While other architectures could also address this problem effectively, this model was chosen due to its simplicity and its proven performance on similar tasks \cite{vgg16_classification}. The model features an input size of (45×90) and produces an output vector of (1×180), followed by a softmax layer. Each output bin represents the probability of the source corresponding to that bin, with a resolution of 2° in the azimuthal angle. This formulation frames the DOA estimation as a classification problem—specifically, azimuth classification—where the network learns to assign likelihoods to discrete angular bins based on spatial features in the input map. This choice was driven by our experimental findings that demonstrated better convergence with this method. Moreover, handling DOA as a classification problem is a common practice in DNN localization algorithms\cite{doa_classification_ones,doa_classification_ones2}.

Mathematically, in SRP-PHAT formulations, azimuth and elevation are inherently coupled through the spatial structure of inter-microphone delays \cite{elevationIfluenceAzimuth1}. For example, with only two microphones, knowing the elevation can constrain the possible azimuth directions to two candidates based on time-delay. While our work does not include a direct study to isolate the contribution of elevation information to azimuth estimation, in this work we have decided to keep elevation information available to the network, similar to recent work that uses joint estimation of azimuth and elevation \cite{Robust_Sound_Source_Tracking_Using_SRP-PHAT_and_3D_Convolutional,SPHERICAL_CONVOLUTIONAL_RECURRENT_NEURAL_NETWORK_FOR_REAL-TIME_SOUND_SOURCE_TRACKING,SurveyDOAnetworks} The focus on azimuth and not elevation stems from the use of a semi-circular array confined to the horizontal plane—representing like headset-mounted microphones.

\subsection{Reliability Estimation}
\label{sec:Reliability_Estimation}
Although existing DNN-based SRP-PHAT methods significantly improve localization accuracy, they typically do not provide measures of prediction reliability, limiting their practical use. To address this, we explicitly integrated reliability estimation into our model by defining a reliability score as the maximum value of the network’s softmax output:

 \begin{equation} C_{\text{DNN}} = \max \left( \text{Softmax}(z_j) \right) \end{equation}

\noindent
where $z_j$	denotes the logit corresponding to the 
$j-th$ azimuth bin. Since the softmax function normalizes the output to form a probability distribution, $C_{\text{DNN}}$ reflects the model’s relative confidence in its predicted direction.

By training the network with Gaussian-weighted labels, we encouraged the model to produce smoother and more calibrated probability outputs, leading directly to meaningful and reliable confidence scores. Furthermore, we systematically studied how varying the Gaussian kernel width during training affects both prediction accuracy and the reliability scores. This enabled us to selectively filter predictions based on their estimated confidence, demonstrating substantial improvements in localization accuracy when using only the most reliable results. Thus, our approach offers a practical method for improving robustness by explicitly training for and utilizing prediction reliability.

\section{Experimental study}
\label{sec:Experimental_study}

\subsection{Dataset}
\label{sec:Dataset}
\begin{figure}[htbp]
    \centering
    \includegraphics[width=0.5\textwidth]{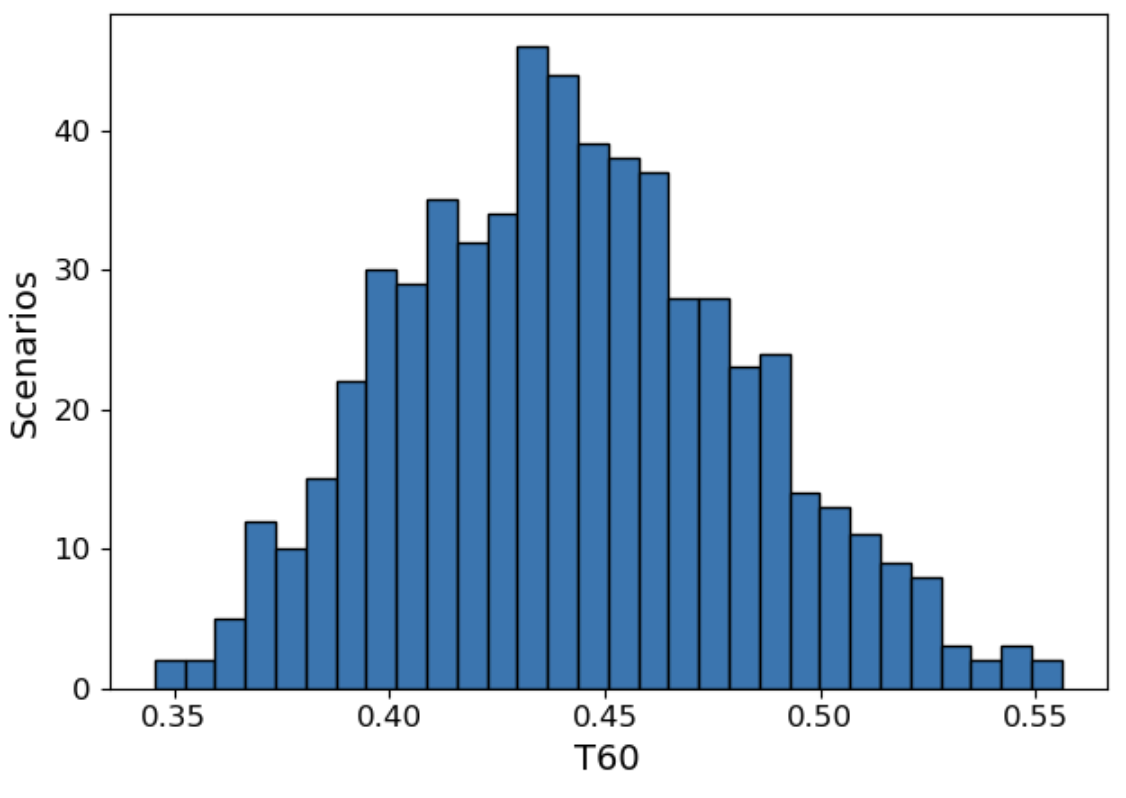}
    \caption{Training data T60 distribution.}
    \label{fig:T60_dist}
\end{figure}

To generate training and evaluation data under controlled yet realistic conditions, we simulated a large corpus of acoustic scenarios using the image method \cite{image_method} for room impulse response generation. Each scenario represents a 2.5-second recording of a speaker in a reverberant room environment, captured by a microphone array. For each scenario, the room dimensions were randomly sampled from a uniform distribution bounded between $[3,5,3]$ m and $[6,10,4]$ m along the (width, length, height) axes. This range was selected to reflect typical indoor environments. The reverberation time (T60) of each room was computed based on its dimensions and absorption coefficients, with the distribution of T60 values shown in Fig. \ref{fig:T60_dist}

The positions of both the speaker and the microphone array were randomized within the room volume. To maintain physically plausible and acoustically relevant setups, we retained only scenarios where the Euclidean distance between the source and array was within $[1/3, 3] \times r_c$, where $r_c$ is the critical distance in the room \cite{criticalDistance}. Additionally, the speaker and the microphone array were positioned on the same horizontal plane, representing normal height of speakers. Having both the array and speakers at the same height is particularly relevant for head-mounted arrays.

\begin{figure}[htbp]
    \centering
    \includegraphics[width=0.4\textwidth]{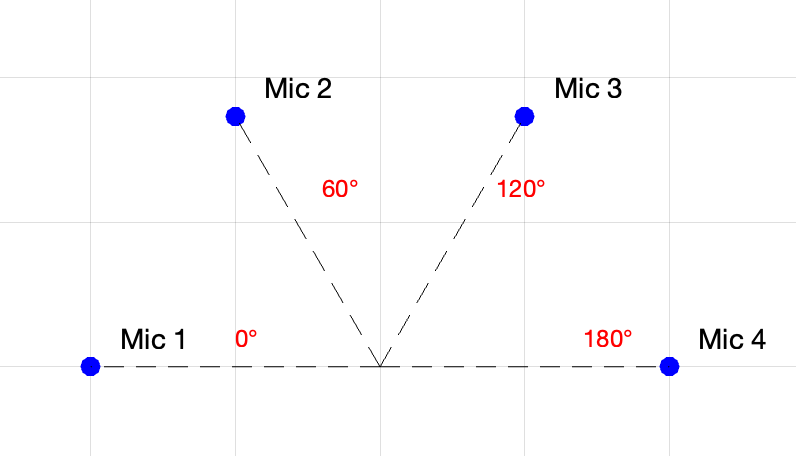}
    \caption{Microphone positions in a semi-circular array configuration. The array consists of four microphones placed at 60° angular intervals around the origin, with a fixed radius of 10 cm.}
    \label{fig:microphonePositions}
\end{figure}

\noindent
For training, speech content was simulated using clean recordings of six different speakers (three male and three female), each producing distinct utterances, from the "Noisy Reverberant Speech Database for Training Speech Enhancement Algorithms and TTS Models" corpus \cite{corpusMixedSpeakers}. For testing, two additional speakers—one male and one female—were used, each providing different utterances not seen during training as well as the simulated rooms. The array configuration used throughout the simulation is a semi-circular microphone array consisting of four evenly spaced microphones placed on the horizontal plane. This geometry closely resembles practical wearable devices such as headset-mounted microphones. The configuration is illustrated in Fig. \ref{fig:microphonePositions}

\subsection{Methodology}
\label{sec:Methodology}

\subsubsection{Model Training}
\label{sec:Model_Training}
To train the proposed deep learning-based DOA estimator, we used the simulated dataset described in Sec. \ref{sec:Dataset}. All audio signals were sampled at 48KHz and segmented into short time frames of 32 milliseconds with 50\% overlap. For each time frame, an SRP-PHAT directional map was computed as detailed in Sec. \ref{sec:srp:extracting_directional_map}. In total, 600 2.5-second long examples were used for training and 100 for evaluation, yielding approximately 82,000 directional maps for training and an additional 13,600 reserved for evaluation.

Each SRP-PHAT map was treated as an input feature to the DNN model, processed as described in \ref{sec:srp:extracting_directional_map}. The corresponding ground-truth DOA was used to construct a target label. While it is common in classification tasks to use one-hot encoded labels \cite{doa_classification_ones,doa_classification_ones2}, typically in combination with cross-entropy loss, this approach treats each azimuth bin as an entirely independent class, ignoring the natural continuity of spatial directions. In this work the labels are smoothed by convolving the one-hot target with a Gaussian kernel (with variance $\sigma^2 = 1$), introducing continuity across neighboring azimuth bins. 
In this work \(\sigma\) denotes the standard deviation of the discrete Gaussian kernel which has array index units and is defined as: 
\begin{equation}
G[k] = \frac{e^{-\frac{k^2}{2\sigma^2}}}{\sum_{j=-K}^{K} e^{-\frac{j^2}{2\sigma^2}}}
\label{eq:normalized_gaussian}
\end{equation}

\noindent
where \(k \in \{-K, \ldots, K\}\) are integer offsets and the radius \(K\) is chosen to be sufficiently large to truncate negligible values. The model was then trained using mean squared error (MSE) loss ,while the variance of the Gaussian kernel selected empirically to balance sharpness and smoothness across neighboring bins. This formulation promotes smoother learning and allows small DOA deviations to be penalized less severely.

\subsubsection{Evaluation Metrics}
\label{Evaluation_metrics}

To evaluate the performance of the proposed DOA estimation model, we employ two complementary metrics: mean absolute error (MAE) and directional accuracy. The MAE measures the average absolute difference, in degrees, between the predicted direction $\hat{\phi}_i$ and the corresponding ground truth direction $\phi_i$
over a total of $N$ time frames:

\begin{equation}
\text{MAE} = \frac{1}{N} \sum_{i=1}^{N} \left| \hat{\phi}_i - \phi_i \right|
\end{equation}

\noindent
 In addition, we define accuracy as the proportion of predictions that fall within a small angular margin 
$\delta = 5^\circ$ from the true direction:

\begin{equation}
\text{Accuracy} = \frac{1}{N} \sum_{i=1}^{N} 1\left( |\hat{\phi}_i - \phi_i| < \delta \right)
\end{equation}

\noindent
where $1(\cdot)$ is the indicator function that equals 1 when the condition is satisfied and 0 otherwise. This reflects the proportion of predictions sufficiently close to the true direction for practical use.

To enable comparison between our method and the classical SRP-PHAT, we defined a reliability score for SRP-PHAT using the following heuristic measure:

\begin{equation} C_{\text{SRP}} = \frac{\max (E(\phi, \theta=90^\circ))}{\frac{1}{M} \sum_{k=1}^{M} E(\phi_k, \theta=90^\circ)}
\label{eq:C_srp}
\end{equation}

\noindent
where $E(\phi,\theta)$ is the steered response energy at $\phi$ and $\theta$ and the denominator is the average energy across all 
$M$ azimuth bins at a fixed elevation angle of 
$90^\circ$, chosen to correspond with the known speaker plane. This score provides a normalized peak-to-average ratio, offering a plausible measure of confidence for SRP-PHAT in our setup.
It is important to note, that unlike the DNN model, $C_{\text{SRP}}$ scoring method leverages prior knowledge that the speaker is always located at 90° elevation, in the same plane as the microphone array. Therefore, the SRP-PHAT method benefits from this constraint, which is not available to the DNN during inference.

\subsection{Results – Overall Performance}
\label{sec:Evaluation}

Evaluation results are presented in Table \ref{tab:eval_results}, where the proposed DNN-SRP-PHAT model clearly outperforms the classic SRP-PHAT. To further investigate the reliability of each method, we defined a certainty score for both the classical and DNN-based approaches.

\begin{table}[h]
\centering
\begin{tabular}{lcc}
\toprule
\textbf{Metric} & \textbf{SRP-PHAT} & \textbf{DNN-SRP-PHAT} \\
\midrule
MAE & 45.9 & 24.9 \\
Accuracy (\%) & 42 & 71.6 \\
\bottomrule
\end{tabular}
\caption{Results of the experimental study for the test set, for the  SRP-PHAT and DNN-SRP-PHAT algorithms.}
\label{tab:eval_results}
\end{table}

\subsection{Effect of Gaussian Label Smoothing}
\label{sec:gauss_smoothing_effect}

To better understand the influence of Gaussian label smoothing on model performance, we conducted a systematic study by training separate models with varying standard deviations $\sigma \in [0.01, 0.2, 1, 1.6, 2.4]$, where the Gaussian kernel was computed as defined in Eq. \ref{eq:normalized_gaussian}. This allowed us to evaluate how the degree of spatial label smoothing affects both the overall accuracy and reliability of the DNN-based DOA estimator.

\begin{figure}[htbp]
    \centering
    \includegraphics[width=0.5\textwidth]{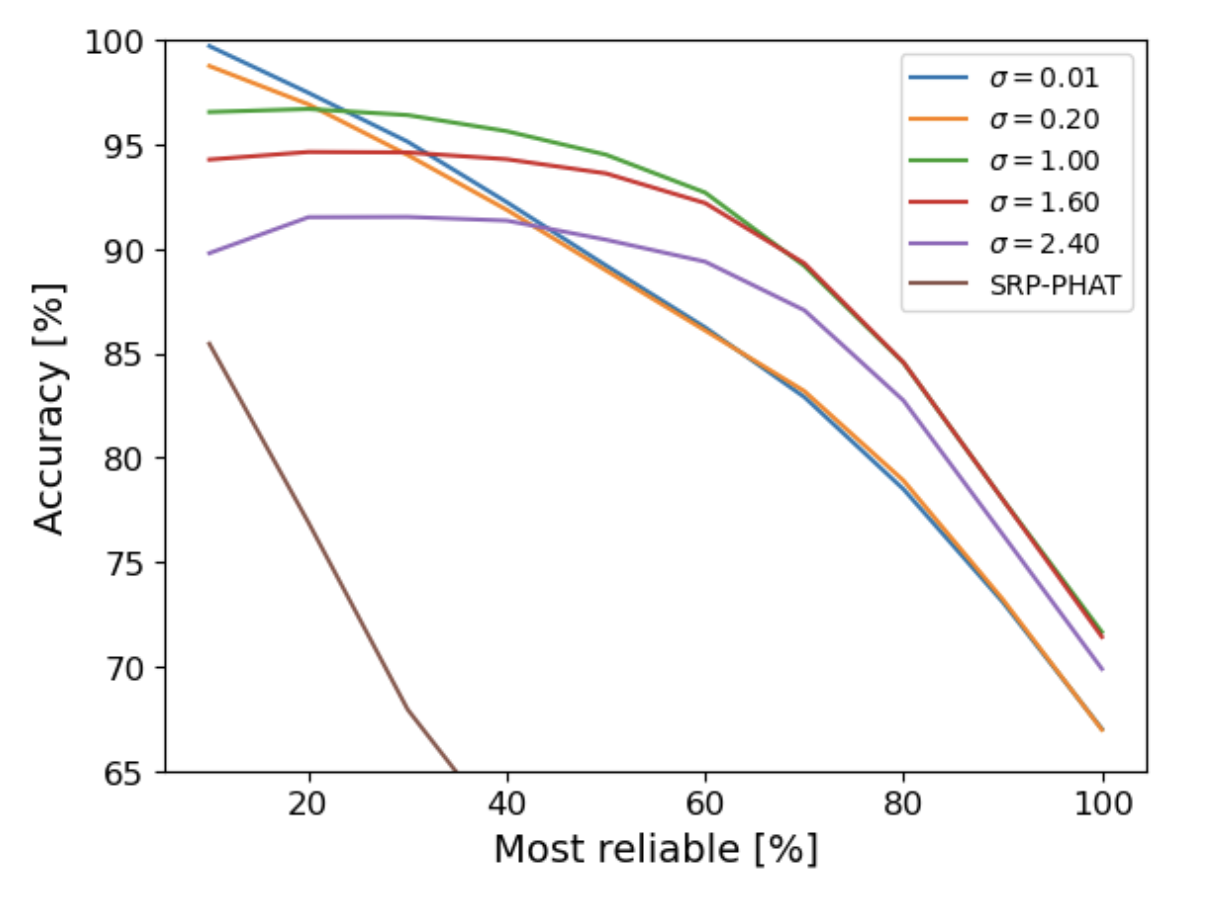}
    \caption{Accuracy as a function of the percentage of most-reliable input frames. Each curve represents a different model trained with a distinct Gaussian kernel used to process the labels.}
    \label{fig:result_accuracyVsTop}
\end{figure}

Table \ref{tab:gaussian_results} presents the results for two evaluation settings: using 100\% of the test data, and using only the top 10\% of time frames as ranked by the model’s confidence score $C_\text{DNN}$. While a particular Gaussian variance (e.g., $\sigma^2 = 1$) yielded optimal results across the full dataset, we observed that the optimal smoothing parameter shifted when considering only the top 10\% of the most confident predictions. This suggests that training with a different Gaussian width may be advantageous when the system is designed to operate in high-certainty regimes. A full comparison is visualized in Fig. \ref{fig:result_accuracyVsTop}. It can be observed that higher values of $\sigma$ yield better accuracy when evaluating 100\% of the data; however, around the 30\% mark, the trend reverses, and lower $\sigma$ values begin to outperform. Notably, in the top 10\% of the data, models trained with lower $\sigma$ achieve a significantly larger performance margin over the others.
Regarding SRP-PHAT, applied using reliability measure as in Eq. \ref{eq:C_srp}, it also shows improved localization accuracy for higher-reliability frames. However, while the classical method demonstrates a more gradual, linear improvement, the DNN model shows a significantly sharper increase in accuracy when lower-certainty frames are excluded.
\begin{table}[h]
\centering
\begin{tabular}{cccccc}
\toprule
\textbf{Std} & \multicolumn{2}{c}{\textbf{Full Dataset}} & \multicolumn{2}{c}{\textbf{Top 10\%}} \\
\cmidrule(lr){2-3} \cmidrule(lr){4-5}
$\sigma$ & MAE & Acc. (\%) & MAE & Acc. (\%) \\
\midrule
0.01 & 29.6 & 67 & 1.5 & \textbf{99.7} \\
0.2 & 28 & 67 & \textbf{1.4} & 98.7 \\
1.0 & 24.9 & \textbf{71.6} & 3.9 & 96.5 \\
1.6 & 24.2 & 71.4 & 7.4 & 94.3 \\
2.4 & \textbf{23.9} & 69.9 & 10.1 & 89.8 \\

\bottomrule
\end{tabular}
\caption{Comparison of model performance under different Gaussian smoothing variances. Accuracy and MAE are reported for all time frames (full dataset) and for top 10\% of frames sorted by reliability score.}
\label{tab:gaussian_results}
\end{table}

\begin{figure}[htbp]
    \centering
    \includegraphics[width=0.5\textwidth]{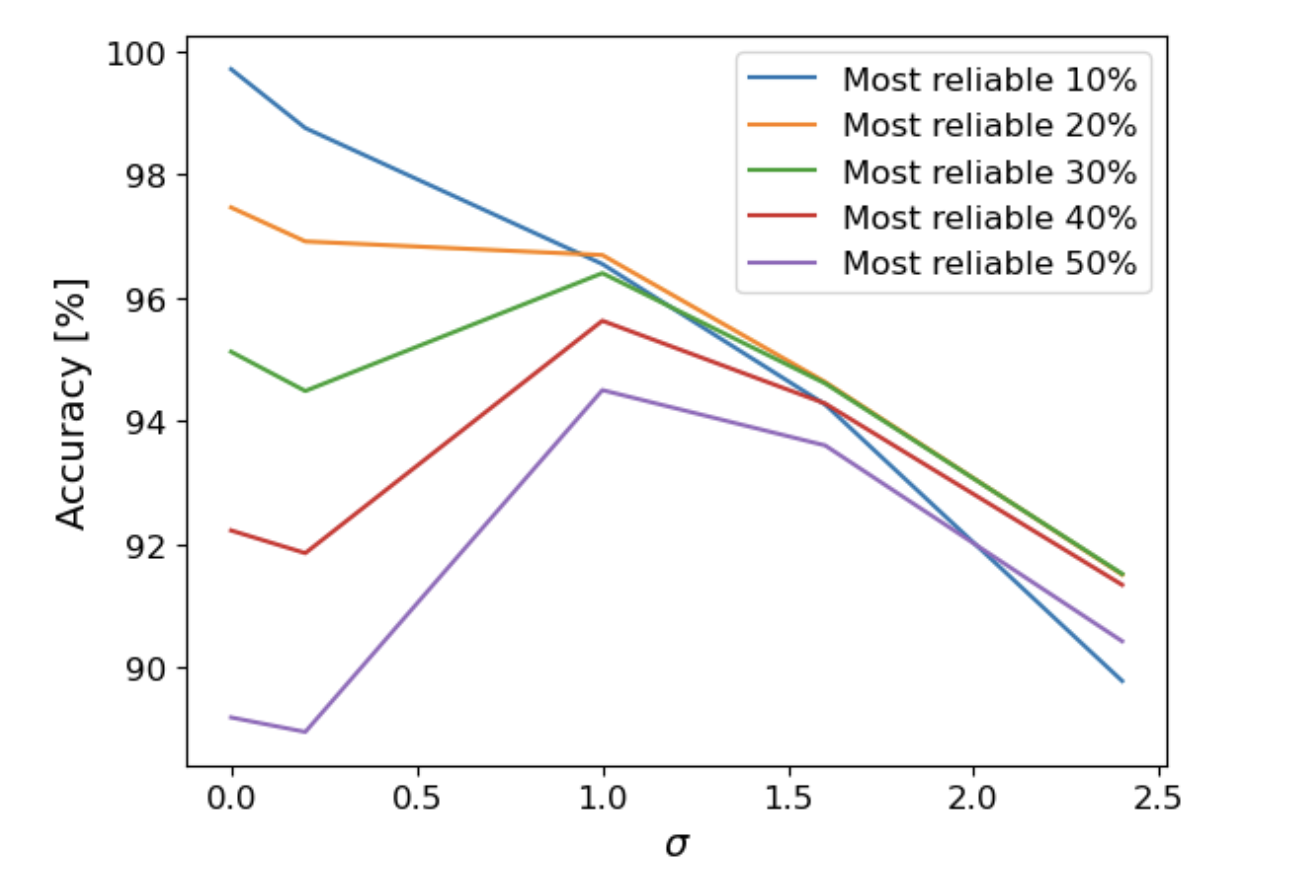}
    \caption{Accuracy as a function of different sigmas used in the training process. Each curve represents the percentage of data that was used sorted by reliability score.}
    \label{fig:result_accuracyVsSigmass}
\end{figure}

\noindent
Fig. \ref{fig:result_accuracyVsSigmass} shows the optimal value of $\sigma$ for each \% of most-reliable input frames, where the optimal $\sigma$ is defined as the one achieving the maximum accuracy for each curve. While $\sigma = 1$ yields better performance when using 30-100\% of the frames, models trained with lower $\sigma$ values perform better when only the top-ranked frames are used.

\noindent
These findings demonstrate the notion that the choice of Gaussian variance in the label construction should be application-aware: models operating in a continuous, high-throughput regime may prefer a wider smoothing, while decision-critical applications that rely on high-confidence frames may benefit from a sharper smoothing configuration.

\section{Conclusion}
\label{sec:conclusion}

This paper proposed and studied a deep learning-based speaker localization method leveraging SRP-PHAT directional maps, specifically designed to output reliable reliability scores. By training our network using Gaussian-weighted labels, we ensured that the model's output provide valid confidence estimates for each prediction. We further demonstrated how explicitly selecting predictions based on these reliability scores significantly improves localization accuracy under challenging acoustic conditions of reverberant speech. Additionally, we systematically investigated how the choice of Gaussian kernel width during training directly influences both prediction accuracy and reliability calibration, highlighting the importance of aligning kernel selection with the reliability demands of specific applications. This carefully designed reliability-aware framework thus provides practical guidance for deploying robust, high-confidence localization solutions in realistic, reverberant environments. Future work may explore extending this approach to multi-speaker localization, dynamic environments, and more principled uncertainty estimation techniques.

%\vfill
%\pagebreak

% \section{REFERENCES}
% \label{sec:refs}


\bibliographystyle{IEEEbib}
% \bibliography{strings,refs}
\bibliography{sample}

\end{document}